\documentclass{article}

\usepackage[UKenglish]{babel}

\usepackage[letterpaper,top=2cm,bottom=2cm,left=3cm,right=3cm,marginparwidth=1.75cm]{geometry}

\usepackage{amsmath}
\usepackage{graphicx}
\usepackage[colorlinks=true, allcolors=blue]{hyperref}
\title{Systematic Uncertainties in Unfolding Considering the Likelihood Formalism}
\author{Lydia Brenner$^1$, Carsten Burgard$^2$ and Vincent Croft$^1$}
\date{%
    $^1$Nikhef -  Dutch National Institute for Subatomic Physics, Amsterdam, The Netherlands\\
    $^2$University of Hamburg - Institute of Experimental Physics, Hamburg, Germany\\[2ex]%
    \today
}
\begin{document}
\maketitle

\begin{abstract}
This paper describes the treatment of systematic uncertainties in a Likelihood formalism. RooUnfold, which includes most of the unfolding methods that are commonly used in particle physics, is used to compare a newly implemented method inside this toolkit to existing methods. The interface with the RooFit statistical software package is used for the treatment of systematic uncertainties. The RooUnfold package with RooFit interface, commonly called RooFitUnfold, provides a common interface to unfolding algorithms as well as common uniform methods to evaluate their performance in terms of bias, variance and coverage. This paper exploits this common interface to compare the performance of unfolding with a Tikhonov regularisation term directly in the likelihood with an unfolding method that optimises a Tikhonov regularisation applied separately of the likelihood formalism. Comparisons are made with and without the treatment of (systematic) uncertainties 
and are applied to an example problem.
\end{abstract}

\section{Introduction}

Unfolding is the solution to an inherently ill-posed problem. The goal of unfolding is the estimate the true physical distributions or parameters from measured data distorted by an imprecise detector. The unfolding process is integral to a wide range of analyses in various science domains. In this paper we focus on high energy physics, where detector resolution, inefficiencies, and acceptance can obscure the underlying kinematic properties of the studied phenomena. Through the statistical estimation of these distortions, unfolding facilitates direct comparisons between experimental results and theoretical predictions, allowing for precise extraction of underlying model parameters. \\


This paper focuses on addressing systematic uncertainties, which we define as uncertainties that are not due to the sample size of the measured data (statistical uncertainties), within the unfolding framework. In particular this paper focuses on the treatment of uncertainties in the likelihood-based formalism. By framing the problem in terms of the underlying probability distributions, we aim to establish a consistent approach for incorporating and propagating both statistical and systematic effects. The paper is organised as follows: section one introduces the mathematical formulation of the unfolding problem and the establishes the formalism of the likelihood construction, section two addresses uncertainties and how statistical models for the unfolding of distributions with uncertainties can be constructed, section three introduces an example distribution and associated representative uncertainties, with various options for unfolding treatments. The paper concludes wwith recommendations for the treatment of statistical and systematic uncertainties in unfolding studies. This paper uses the in-Likelihood implementation in RooUnfold \cite{1910.14654}, another implementation of the in-likelihood method of unfolding can be found in the CMS combine tool \cite{cmscollaboration2024cmsstatisticalanalysiscombination}.

\subsection{Definition of the unfolding problem}
\label{sec:definition}
For the definition of the unfolding problem we follow the formulation, notation and terminology of Ref.~\cite{1910.14654} precisely. A short summary of the necessary notation and definitions are given here. While the unfolding principle can be applied to distributions in general, this paper focusses on a counting experiment where a histogram of some variable $x$ is used to construct a histogram.\\ 

The expected number of events in $M$ bins that would be found if the variable $x$ could be measured exactly for each event is defined as $\vec{\mu} = (\mu_1, \ldots, \mu_M)$. This is called this the \textbf{``true histogram''}.
The measured values used to construct a histogram in $N$ bins are defined as $\vec{n} = (n_1, \ldots, n_N)$. This is called the as the \textbf{``data histogram''} (or simply ``the data'')'. These values will differ stochastically from the true value by some amount due to the limited resolution of the detector.
The data values follow some probability distribution and have expectation values $\vec{\nu} = E[\vec{n}] = (\nu_1, \ldots, \nu_N)$.  This is called the \textbf{``expected data histogram''}.\\

The goal of unfolding is to estimate the parameters of the ``true histogram'', using the ``data histogram''. Estimators are denoted with a circumflex, i.e., $\hat{\vec{\mu}} = (\hat{\mu}_1, \ldots, \hat{\mu}_M)$.
With a perfect detector the values of $\vec{\mu}$, $\vec{\nu}$ and $\vec{n}$ are trivially related, however with an imperfect detector a true value of the variable $x$ in a given bin may be measured in a different one.  The migration of events is expressed as

\begin{equation}
\label{eq:nuRmu}
\nu_i = \sum_{j=1}^M R_{ij} \mu_j \;,
\end{equation}

where $i = 1, \ldots, N$, and $R_{ij}$ is referred to as the \textbf{response matrix} defined as,

\begin{equation}
\label{eq:responsemat}
R_{ij} = P(\mbox{measured value in bin } i|\mbox{true value in bin }j) \;,
\end{equation}

\noindent gives the conditional probability for an event to be
reconstructed in bin $i$ of the data histogram given that its true was
in bin $j$ of the true histogram.\\



The likelihood function is defined as $\mathcal{L}(\vec{\mu}) = P(\vec{n} | \vec{\mu})$. Assuming the $n_i$ can be modelled as independent and Poisson distributions, the likelihood is

\begin{equation}
  \label{eq:poissonlnl}
  \mathcal{L}(\vec{\mu}) = \prod_{i=1}^N \frac{\nu_i^{n_i}}{n_i!} e^{- \nu_i} \;,
\end{equation}

where $\nu_i = \sum_{j=1}^M R_{ij} \mu_j$.  

Assuming the observed number of events can be modelled as following a Gaussian distribution with mean $\nu_i$ and standard deviation $\sigma_{n_i}$, the log-likelihood is (up to an additive constant) a sum
of squares,

\begin{equation}
\label{eq:gausslnl}
  \ln \mathcal{L}(\vec{\mu}) = - \frac{1}{2} \sum_{i=1}^N \frac{(n_i -
    \nu_i)^2}{\sigma_{n_i}^2} \;.
\end{equation}

While the standard deviation $\sigma_{n_i}$ encompasses the statistical uncertainty on the data histogram. While this has a clear definition for the inclusion of the statistical uncertainty of the data histogram, other uncertainties still need to be included. How this can be done will be discussed in section \ref{sec:systematics}. 

\subsection{A likelihood based solution to the unfolding problem}
\label{sec:unfoldingmethods}
In the case of the same number of bins for $\vec{\mu}$ and $\vec{\nu}$, meaning $N = M$ and the response matrix $R$ is square, and if $R$ is
nonsingular\footnote{Nonsingular is not a strong assumption since realistic matrices are typically diagonally dominated and thus nonsingular.} we can relate the $\vec{\mu}$ and $\vec{\nu}$ vectors as $\vec{\mu} = R^{-1} \vec{\nu}$. The estimators for $\vec{\mu}$ could then be calculated as

\begin{equation}
\label{eq:matrixinv}
\hat{\vec{\mu}} = R^{-1} \vec{n} \;.
\end{equation}

In the likelihood as described in Eq. \ref{eq:poissonlnl} this corresponds to the maximum-likelihood estimators for the $\nu_i$ being $\hat{\nu}_i = n_i$. \\

Without the requirement of the same number of bins for the two histograms, using the likelihood formalism introduced in Eq. \ref{eq:gausslnl}, the estimators $\hat{\vec{\mu}}$ can be chosen to correspond to the maximum of the log-likelihood $\ln \mathcal{L}_{\rm max}$. However, the maximum of the log-likelihood could have a large variance, caused by the inherent  assumption of this choice that any data fluctuations are due to the underlying model rather than statistical fluctuations. \\
To solve this problem, a $\hat{\vec{\mu}}$ can be chosen that does not correspond to the maximum of the log-likelihood $\ln \mathcal{L}_{\rm max}$, but rather one considers a region of $\vec{\mu}$-space where $\ln \mathcal{L}(\vec{\mu})$ is within some threshold below $\ln \mathcal{L}_{\rm max}$. The optimal estimators $\hat{\vec{\mu}}$ can then be chosen to correspond to the distribution that is smoothest by some measure, which negates the assumption of statistical fluctuations comes from the underlying model. This can be achieved by maximising not $\ln \mathcal{L}(\vec{\mu})$ but rather a linear combination of it and a regularisation function $S(\vec{\mu})$, which represents the smoothness of the histogram $\vec{\mu}$.  That is, the estimators $\hat{\vec{\mu}}$ are determined by the maximum of

\begin{equation}
  \label{eq:regular}
  \varphi(\vec{\mu}) = \ln \mathcal{L}(\vec{\mu}) +  \tau S(\vec{\mu}) \;,
\end{equation}

where $\varphi(\vec{\mu})$ replaces the role of the $\ln \mathcal{L}(\vec{\mu})$ term in an analysis, and the regularisation parameter $\tau$ determines the
balance between the two terms and therefore sets the measure of smoothness.  \\

\noindent When inserting the likelihood from equation \ref{eq:gausslnl} this gives

\begin{equation}
\label{eq:gausslnlreg}
  \varphi(\vec{\mu}) = 
  - \frac{1}{2} \sum_{i=1}^N \frac{(n_i -
    \nu_i(\vec{\mu}))^2}{\sigma_{n_i}^2} + \tau S(\vec{\mu}).
\end{equation}

The regularisation function $S(\vec{\mu})$ can be chosen in a number of ways. One commonly used function is Tikhonov regularisation \cite{Tikhonov77}. For discrete bins $\vec{\mu}$, the Tikhonov regularisation function, which is used for the remainder of this paper, can be expressed as

\begin{equation}
\label{finite_diff2}
S(\vec{\mu}) = - \, \sum_{i=1}^{M-2} (-\mu_{i} \, 
+ \, 2 \mu_{i+1} \, - \,
\mu_{i+2})^{2} .
\end{equation}







\section{Treatment of systematic uncertainties}
\label{sec:systematics}

Systematic uncertainties play a critical role in any analysis involving unfolding and must be properly accounted for within the likelihood framework \cite{Conway:2011in}. These uncertainties are incorporated as nuisance parameters that modify the expected bin counts, $\nu_i$, and are constrained by auxiliary terms in the log-likelihood. These additional terms account for the deviations introduced by systematic effects and ensure that their contributions are coherently propagated through the unfolding procedure.

As described in section \ref{sec:definition}, the data as measured by the detector can be considered statistically independent for each bin and a Poisson error can be assigned based on the bin count in each bin assuming no correlations between bins. This is stated as the parameter $\sigma_{n_i}$. This uncertainty, and only this, is considered a \textbf{statistical uncertainty.} Typically $R_ij$ is constructed from synthetic models of the data sampled from the joint distribution of truth and expected events using Monte Carlo (MC) sampling techniques. For the MC created for the study, the same procedure can be applied of Poisson bin-counts, however this can be considered as a \textbf{systematic uncertainty} since it is not directly related to the taken data with the detector, but rather based on theoretical predictions and the user has some control over the precision of the created samples.\\

MC statistical uncertainties arise due to the finite size of the simulated samples used to model the detector response and background contributions. These uncertainties are modelled by introducing nuisance parameters that scale the contributions in each bin of the simulation. For instance, a parameter $\gamma_{\text{MC},i}$ may be introduced to account for fluctuations in the expected counts derived from the simulation, leading to modified terms in $\nu_i$. These nuisance parameters are typically constrained by independent Poisson distributions reflecting the finite statistics of the simulated data, and their contribution to the log-likelihood takes the form of independent terms added to the likelihood introduced in Equation \ref{eq:gausslnl}.\\

Background contributions introduce another source of uncertainty, particularly when their magnitude or shape is not precisely known. Historically, these contributions are often subtracted prior to unfolding. However, since these contributions are systematically dependent on many of the same parameters as the target `signal' component, a more correct procedure in the case of estimating the effect of these uncertainties with respect to the data, is of course to estimate the background contributions together in combination with those of the signal. These contributions are modelled in the expected bin counts, $\nu_i$, through a parametrized background function, in which the shape parameters $\beta$ are used to estimate the affect of \emph{known} effects on the shape of the MC modelling with the background estimation. 

Scaling factors for the overall normalization of the background or signal parameters, such as those related to integrated luminosity or theoretical cross-section predictions, affect the overall scale of the expected contributions. Each of these contributions is modelled by a single nuisance parameter, denoted here as $\alpha$, which uniformly scales each component of the likelihood (such as the signal or background) by some factor across all bins in the likelihood. Normalization uncertainties are generally independent of shape uncertainties but can be correlated with other normalization parameters, depending on the analysis. 

A final distinct source of uncertainty arises from the unfolding procedure itself, particularly when relating to the regularization applied to stabilize the solution. The choice of $S(\vec{\mu})$ introduces a methodological uncertainty that depends on the regularization form, such as penalizing large gradients or enforcing curvature constraints. This term, parametrized by $\upsilon$, can introduce correlations between bins in $\vec{\mu}$, directly influencing the unfolded result and its uncertainty.

These combine to augment equation \ref{eq:gausslnlreg} to become:
\begin{equation}
\label{eq:gausslnlregsys}
  \varphi(\vec{\mu},\alpha,\beta,\gamma,\theta) = 
  - \frac{1}{2} \sum_{i=1}^N \frac{(n_i -
    \nu_i(\vec{\mu},\alpha,\beta,\gamma))^2}{\sigma_{n_i}^2} + \tau S(\vec{\mu},\theta).
\end{equation}
Where $\alpha$ is a parameter governing the normalization of all MC samples, and as such is uniform over bins $i$; $\beta$ relates the different bins in $i$ to account for variations in the MC shape; $\gamma$ is a parameter to determine the effect of the MC statistical uncertainty (usually a Gaussian scaling factor over all components contributing to $i$); and $\theta$ are parameters that specifically effect the smoothing. The regularisation strength $\tau$ on the other hand is not considered a variable parameter, but rather a fixed parameter governing the overall scale of the smoothing component, which needs to be chosen appropriately as discussed in the following section.

When combined with the systematic effects already discussed, such as those affecting the response matrix, background contributions, and normalization, the regularization term completes the statistical model. The total likelihood framework ensures that these contributions, including the unfolding uncertainties, are coherently treated and propagated into the final unfolded distribution, allowing for a consistent estimation of $\vec{\mu}$ and its associated uncertainties.\\

For all methods included in RooUnfold, other than the in-likelihood method introduced in this paper, systematic effects are propagated through an explicit covariance construction. Each source of uncertainty is represented by a corresponding contribution to the total covariance matrix of the reconstructed spectrum. The statistical component is derived from the Poisson variance of the nominal bin counts. For each systematic variation, both the upward and downward shifts are evaluated at the 1-sigma-level and compared to the nominal prediction and added in quadrature to obtain the variance contribution per bin. These components are accumulated into a single covariance matrix that reflects the combined uncertainty on the detector-level signal and background templates. This matrix is then be propagated through the unfolding response.\\
For the in-likelihood method, the systematic effects are implemented directly as constraint terms in the likelihood itself.

\subsection{Bias, variance and coverage}
\label{sec:bvc}
\noindent One of the most important properties of the estimators that we investigate in this paper is the \textbf{bias}. The bias is defined as the difference between estimator's expectation, $E[\mu]$, and
parameter's true values, $\mu$. In each bin $i$ this can be fined as

\begin{equation}
\label{eq:bias}
b_i = E[\hat{\mu}_i] - \mu_i \;,
\end{equation}

\noindent In addition to the bias this paper considers the the covariance matrix $\mbox{cov}[\hat{\mu}_i,
  \hat{\mu}_j]$, in particular its diagonal elements, \textbf{the variances}
$V[\hat{\mu}_i]$ (also written $\sigma_{\hat{\mu}_i}^2$).  In the
  Poisson case with a square nonsingular response matrix, the
  maximum-likelihood estimators have zero bias and their covariance
  equals the minimum variance bound (see, e.g., Ref.~\cite{cowan1998statistical}).
  In the region where the bias is minimised, the variances are extremely large and in this sense the
  unfolding problem is said to be ill-posed
  \cite{Hansen:1990:DPC:98694.98703}.  In unfolding one therefore
  attempts to reduce the variance through some form of regularisation,
  and this necessarily introduces some bias.\\

\noindent The bias can be estimated by performing the unfolding on a synthetic dataset created at the reconstruction level. This dataset follows the description of an "Asimov" dataset\cite{1007.1727}, defined as a Monte Carlo dataset created with the properties of the measured data that behaved according to the nominal predictions from theory. \\
To calculate the bias more accurately, we perform the following procedure based on pseudo-experiments. Each pseudo-experiment (referred to as a 'toy') is a simulated dataset generated with random fluctuations around the theoretical prediction.
First the uncertainties are taken truth level from the unfolded Asimov dataset. A first generation of toys, called \textit{level 1 toys}, are drawn from a distribution around the Asimov truth values based on the known variances in each bin as defined by the Poisson distribution of the expected number of events. Based on the prediction a second generation of toys is drawn, called \textit{level 2 toys}, with the truth values of the level 1 toy acting as central values.
Each of the level 2 toys is folded and then unfolded with the chosen unfolding method using the original response matrix. The bias for each level 2 toy is calculated as 
\begin{equation}
\mathrm{bias}_{l2}=(\sigma_{\mathrm{refold}}-\sigma_{\mathrm{truth}})/\sigma_{\mathrm{truth}}, 
\end{equation}
where the $truth$ refers to the value of the level 1 toy at truth level from which the level 2 toy is thrown and $refold$ refers to the value of the level 2 toy after folding and unfolding. The bias of each bin in the distribution that is unfolded is the average over all $bias_{l2}$. \\
Since the bias is used to estimate the shape effects introduced by the unfolding, the used Asimov dataset should be created using the measured total number of events rather than the predicted. An Asimov dataset with the predicted total number of events can be used to estimate the size of the bias before the measurement is performed. \\

\noindent The \textbf{coverage} is the probability that the true value of $\mu_i$ falls between plus and minus one standard deviation about the estimator of $\mu_i$. It is important to realise that the bias, variance and coverage are always defined with respect to the chosen true and expected data histograms.

\section{Studies on example distributions}
While realistic physics models and detector response functions in HEP rely on elaborate MC simulation packages that simulate both the physics and
detector response from first principles, we opt in this study for simple and analytically expressed physics functions and resolution models. These analytical models represent a 
more portable benchmark than the elaborate simulation models, while still introducing a realistic level of complexity. For the studies done in this paper we use RooFitUnfold. The RooFitUnfold framework provides a unified software environment in which various unfolding algorithms used in HEP are implemented. \\

\subsection{Distribution definition}
\begin{figure}[!ht]
\noindent \begin{centering}
\includegraphics[width=.49\textwidth]{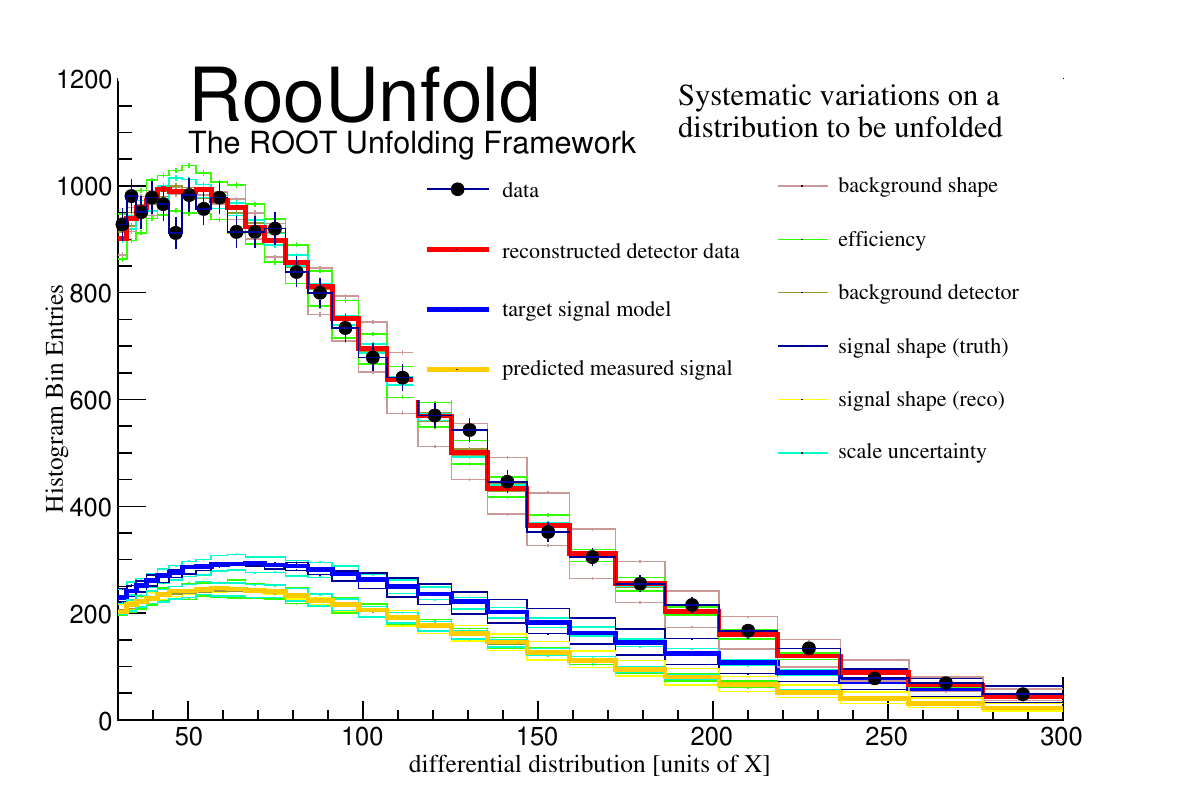}
\includegraphics[width=.49\textwidth]{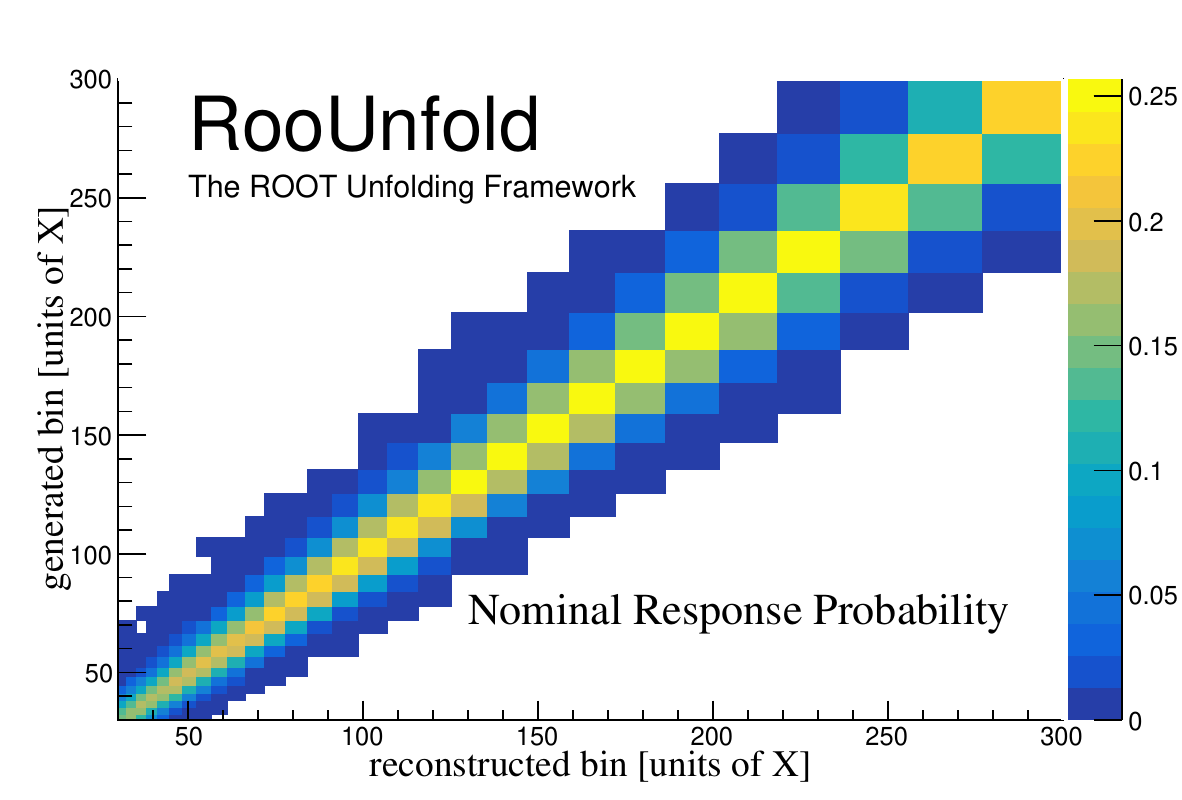}
\par\end{centering}
\caption{Left: The true distribution for the signal plus background model and the corresponding smeared dataset and a simulated measured dataset drawn from the same distribution as the smeared dataset. Right: the response matrix for the signal plus background model, which is populated by the same events as the true distribution shown.}
\label{fig:model_exp}
\end{figure}

We define as benchmark model an exponential decay distribution, smeared with a resolution function that is loosely inspired on a calorimeter response:

\begin{eqnarray*}
f(x|\alpha) & = & f_\textrm{physics}(x_\textrm{true}|\alpha) \ast f_\textrm{detector}(x_\textrm{true},x) \\
            & = & \left(\exp(-a  \cdot x_\textrm{true}) \right) \ast b  \cdot  \textrm{Gauss}\left(0, c \cdot \sqrt{x_\mathrm{true}}/d\right),
\end{eqnarray*}

\noindent where the $\ast$ symbol represents the convolution operator. We define two models variants, labeled signal and background, that correspond to an exponential distribution with a slope $a$ of 0.1 and 0.09 respectively. The $b$ parameter represents a luminosity scale factor and is set to 1. The $c$ and $d$ parameters in this case are set to 1 and represent the detector resolution. Finally a detector efficiency is introduced by
\begin{equation}
x_{\mathrm{eff}} = e - |x_\mathrm{true}|/600
\end{equation}
where an event is thrown out if $x>x_{\mathrm{eff}}$ with $x$ a random number between 0 and 1.  
The true and expected distributions $\vec{\mu}$ and $\vec{\nu}$ corresponding to this model are defined by bins in the range $[0,300]$. 
The true distribution and data for signal and background, populated with 1 million events each, are shown in Fig.~\ref{fig:model_exp}, along with the response matrix.\\

\noindent To get a realistic example of an unfolding problem the \textit{reconstructed detector data} and the \textit{predicted measured signal} are used to fill the response matrix, while the \textit{data} points are created as a statistical fluctuation around the simulated \textit{reconstructed detector data}.  

\subsection{Uncertainties considered}
We assume that the data as measured by the detector can be considered statistically independent and a Poisson error can be assigned based on the bin count in each bin assuming no correlations between bins. We consider this, and only this uncertainty, a \textbf{statistical uncertainty.} \\\\
For the MC created for the study, the same procedure can be applied of Poisson bin-counts, however we consider this a systematic uncertainty since the user has some control of the number of MC events created. \\\\
All other uncertainties that have something to do with the measurement are considered \textbf{systematic uncertainties}. This could, for example, entail the resolution of the detector, an efficiency in identifying a type of process, or an uncertainty in the duration of the measurement. These types of uncertainty could effect both the shape and normalisation of (part of) the measured spectrum. These are often included as a Gaussian constraint in the Likelihood, although other types of constraints do exists depending on the measurement. \\\\
Finally, there are uncertainties coming from the measurement methodology itself, such as the unfolding procedure, in this case referred to as \textbf{unfolding uncertainties}. Here you can consider for example the precision of the Response matrix. This uncertainty can be led back to the limited MC data size in that case. \\\\ 
Collectively, all uncertainties that are not statistical uncertainties, most importantly the systematic uncertainties, are considered as \textbf{nuisance parameters} in the likelihood. In particular this means that we can re-parametrise the likelihood as
\begin{equation}
 L(\vec{\mu}) \rightarrow L(\vec{\mu}|\vec{\theta})
\end{equation}
where $\vec{\theta}$ describe the full set of these nuisance parameters. \\\\

\noindent For the purpose of the study in this paper several different nuisance parameter sets are considered.

\begin{itemize}
    \item Nuisance parameter 1 affecting a scale factor per process 
    \item Nuisance parameter 2 affecting the smearing function 
    \item Nuisance parameter 3 affecting the measurement efficiency 
    \item Nuisance parameter 4 affecting the background curve shape 
    \item Nuisance parameter 5 affecting the signal curve shape 
    \item Nuisance parameter 6 affecting the total scale of the dataset (e.g. Luminosity)
\end{itemize}

These six nuisance parameters are representative of the dominant systematic uncertainties encountered in typical particle physics measurements. Normalization uncertainties (NP1, NP6) correspond to overall scale factors such as luminosity determination, trigger efficiencies, and theoretical cross-section uncertainties—these affect event yields uniformly across the kinematic phase space. Detector response systematics (NP2, NP3) model instrumental effects: energy or momentum resolution smearing (NP2) represents uncertainties in the detector's ability to precisely measure particle energies, analogous to jet energy resolution (JER) or calorimeter response uncertainties in collider experiments, while position-dependent efficiency variations (NP3) capture geometric acceptance, reconstruction efficiency fall-offs at detector boundaries, or particle identification efficiency as a function of kinematic variables. Shape uncertainties for both signal and background processes (NP4, NP5) arise from theoretical modelling choices—such as parton distribution function (PDF) variations, parton shower and hadronisation model differences (comparing Pythia versus Herwig generators), initial and final state radiation (ISR/FSR) modelling, or control region extrapolation uncertainties in data-driven background estimates. Together, these parameters span the principal categories of systematic effects: overall normalization, resolution smearing, acceptance losses, and process shape modelling. The specific implementations and their effects on the distributions are detailed in appendix \ref{app:nps}, demonstrating how each uncertainty manifests in both the response matrix and the reconstructed distributions

\subsection{Tikhonov regularised methods}
As shown in Equation \ref{finite_diff2} a Tikhonov regularisation term can be added to the Likelihood. To compare the in-likelihood implementation to traditional binned methods, the best comparison can be made with methods that also reply on Tikhonov regularisation such as SVD and TUnfold. A short description of these two methods are described below, before a comparision with in-likelihood unfolding is made. The comparison is made both with and without systematic unceratainties. Data statistical uncertainties are always included. 

\subsubsection*{SVD - Singular Value Decomposition}
RooUnfoldSvd provides an interface to the TSVDUnfold class implemented in ROOT by Tackmann \cite{tackmann},
which uses the method of H\"ocker and Kartvelishvili \cite{hocker}.
Singular Value Decomposition is used to express the detector response as a linear series of coefficients.
Regularisation is applied in the form of a truncation of this series in order to remove contributions with a small singular value, $s^2_i \rightarrow s^2_i/(s^2_i + s^2_k)$, where the $k$th singular value defines the cut-off, which correspond to high-frequency fluctuations.


\subsubsection*{In-likelihood regularisation term}
By minimising the $\ln L(\vec{\mu})$ term and $\tau S(\vec{\mu})$ terms in Equation \ref{eq:gausslnlreg} separately, the estimate of $\mu$, e.g. $\hat{\mu}$, is not guaranteed to be the same for the two terms. However, if the estimation of the two terms can happen simultaneously, the two terms would automatically have the same estimator. 
Studying Equation \ref{eq:gausslnlreg}, Equation \ref{finite_diff2} and Equation \ref{eq:gausslnlregsys}, it can be combined to give

\begin{equation}
\label{eq:combinedL1}
 \varphi(\vec{\mu},\alpha,\beta,\gamma,\theta) = 
  - \frac{1}{2} \sum_{i=1}^N \frac{(n_i -
    \nu_i(\vec{\mu},\alpha,\beta,\gamma,\theta))^2}{\sigma_{n_i}^2} - \tau \, \sum_{i=1}^{M-2} (-\mu_{i} \, 
+ \, 2 \mu_{i+1} \, - \,
\mu_{i+2})^{2}.
\end{equation}
Which shown the way the Tikhonov regularisation term smooths the difference between the neighbouring bins. 
However, by comparing to Equation \ref{eq:regular} it is clear that this adjusts the $\ln L(\vec{\mu})$ rather than the likelihood itself. Therefore, the term that can be added to the likelihood itself is the exponential of the regularisation term introduced in Equation \ref{finite_diff2};
\begin{equation}
\label{eq:combinedL2}
 \varphi(\vec{\mu},\alpha,\beta,\gamma,\theta) = 
  \ln (L(\vec{\mu,\alpha,\beta,\gamma,\theta}) +e^{- \tau \, \sum_{i=1}^{M-2} (-\mu_{i} \, 
+ \, 2 \mu_{i+1} \, - \,
\mu_{i+2})^{2}}).
\end{equation}

This allows for simultaneous estimation of the parameters of interest $\hat{\mu}$ and unfolding, taking all uncertainties into account. A key advantage of this approach is that by including systematic uncertainties as nuisance parameters directly in the likelihood, the profile likelihood framework guarantees proper frequentist coverage of confidence intervals (assuming regularity conditions), independent of the regularisation strength $\tau$. Traditional methods that apply systematics corrections post-unfolding require explicit coverage validation for each regularisation choice.\\

When dealing with the systematics uncertainties included in Equation \ref{eq:combinedL2}, in practice the user can  multiply the response matrix with the covariance matrix of the minimised $\ln L(\vec{\mu})$, and with that get an accurate estimation of the total uncertainty on the unfolded distribution. By estimating $\hat{\mu}$ for the regularisation term and the likelihood term simultaneously, it is possible to not only take the variance into account due to the statistics of the data sample during the choice of regularisation strength but the whole covariance with systematics uncertainties incorporated.

\subsection{Choice of regularisation strength}
To make a fair comparison, the regularisation strength for each method is chosen such that it unconditionally minimises the bin-averaged mean squared error (MSE), which corresponds to the sum of bias squared and variance, while still having coverage as explained in Ref.~\cite{1910.14654}. The definitions of bias, variance and coverage are explained in section \ref{sec:bvc}. The choice of regularisation strength by optimising the MSE while requiring coverage for each of the methods is shown in Table \ref{tab:reg}.\\

\begin{table}[h!]
    \centering
    \begin{tabular}{c||c|c}
    Method & Without systematics & With systematics\\
    \hline
    Number of SVD components  & 2  & 11 \\
    in-likelihood $\tau$ & 11   & 3
    \end{tabular}
    \caption{The choice of regularisation strength by optimising the MSE while requiring coverage for each of the methods in the comparison}
    \label{tab:reg}
\end{table}

\subsection{Results}
\begin{figure}[!ht]
\noindent \begin{centering}
\includegraphics[width=.49\textwidth]{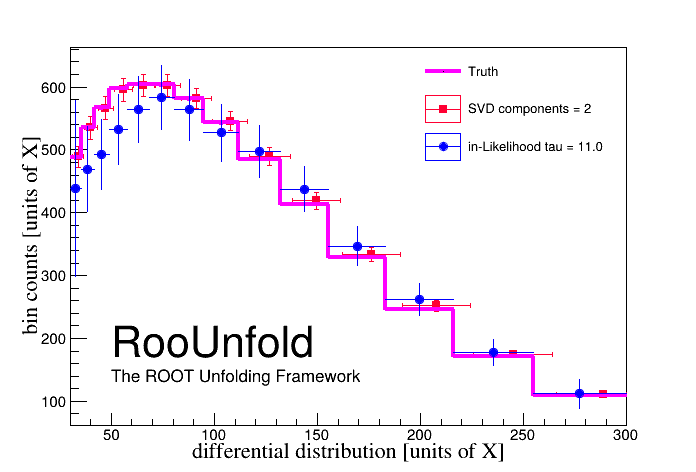}
\includegraphics[width=.49\textwidth]{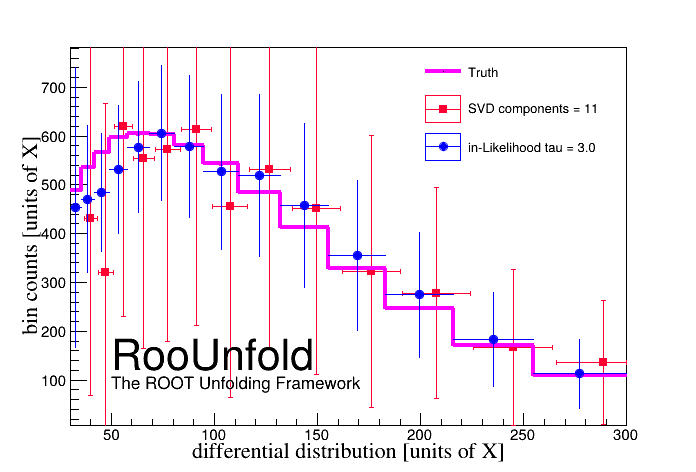}
\par\end{centering}
\caption{Left: Unfolded distribution without systematic uncertainties. Right: Unfolded distribution with systematic uncertainties. For each case the MSE has been minimised, while requiring coverage, to set the appropriate regularisation strength. Due to the fluctuation of the data around the prediction, a perfect match of the unfolded histogram with the Truth distribution is not expected. }
\label{fig:comparison}
\end{figure}

As described above, the \textit{data} used for the comparison is created as a statistical fluctuation around the \textit{reconstructed detector data}. If the \textit{reconstructed detector data} would be unfolded you would expect perfect closure with the \textit{truth} distribution, while you would expect some fluctuations around the \textit{truth} for unfolding the \textit{data} distribution shown in Figure \ref{fig:model_exp}. \\
Two comparisons are made between the SVD method and the in-likelihood method introduced in this paper. The first comparison is made for the case where you are only considering statistical fluctuations, the second comparison includes the systematic uncertainties introduced in section \ref{sec:systematics}. Both comparisons are shown in Figure \ref{fig:comparison}.\\
In the case of a comparison without systematics (Figure \ref{fig:comparison} left) it is clear that the variance of SVD is smaller than that of the in-likelihood method, while on the other hand a stronger bias towards the expected Truth distribution is found since, as explained above, a perfect match with the Truth distribution is not expected in this case. In the case of a comparison including systematics (Figure \ref{fig:comparison} right) you can see that in-Llkelihood has both a smaller variance and smaller bias than the SVD method. Both methods use Tikhonov regularisation, so the main difference comes from simultaneously estimating $\hat{\mu}$ for the $\ln L(\vec{\mu})$ term and $\tau S(\vec{\mu})$ term. 

\begin{figure}[!ht]
\noindent \begin{centering}
\includegraphics[width=.49\textwidth]{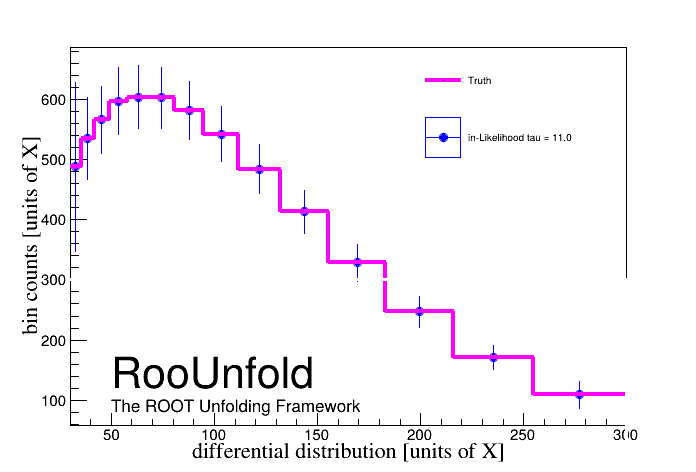}
\includegraphics[width=.49\textwidth]{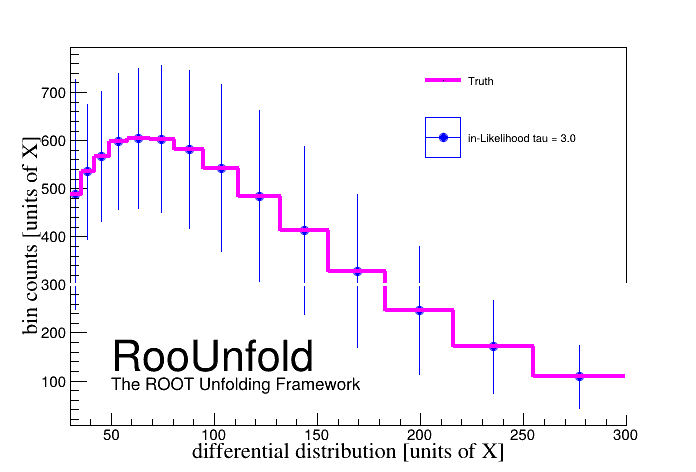}
\par\end{centering}
\caption{Left: Unfolded distribution without systematic uncertainties. Right: Unfolded distribution with systematic uncertainties. Using the same regularisation strength as in Figure \ref{fig:comparison}. Here a perfect match of the unfolded histogram with the Truth distribution is expected. }
\label{fig:comparison_truth}
\end{figure}

\noindent As a cross-check, the unfolded distribution using the in-likelihood method is compared to the truth distribution for the \textit{reconstructed detector data}, where perfect closure between the two distributions is expected, using the same regularisation strengths as above. This check is shown in Figure \ref{fig:comparison_truth}.

\section{Conclusion and outlook}
By framing the problem in terms of the underlying probability distributions, it is possible establish a consistent approach for incorporating and propagating both statistical and systematic effects. It allows to simultaneously estimate the parameters of interest $\hat{\mu}$. The in-likelihood method trivially extends to a multi-dimensional case, where only the regularisation term $S(\vec{\mu})$ needs to be adjusted. In the same way it can be extended for an unbinned case. For visualisation purposes this paper focusses on a binned, one-dimensional use-case. \\\\
While this paper focusses on a Tikhonov regularisation term in the likelihood, other types of regularisations can be included instead. The choice of this term for this paper was for comparison reasons with at least one other method included in RooUnfold. The in-likelihood method, independent on the exact form of the regularisation term, allows for systematics to no longer be evaluated in-situ, while every method included in RooUnfold only includes an approximation of the MLE in presence of systematics. Furthermore, the in-likelihood method guarantees coverage for every value of the regularisation strength inherently from the method, which simplifies the choice of regularisation strength for the user.\footnote{In practice the coverage can be seen as an increase in the variance in the region where for other methods coverage would not have been achieved.} \\\\
This paper introduces the in-likelihood method as incorporated in the RooUnfold package, and demonstrates how systematics can be treated inclusively in the method rather than evaluated in-situ. In addition these results also show a reduction in both the bias and variance that arises naturally as a consequence of this construction when compared to other unfolding methods that also include a Tikhonov style regularisation term.


\bibliographystyle{alpha}
\bibliography{main}

\newpage
\appendix

\section{Appendix: Nuisance parameters}
\label{app:nps}
This appendix provides technical details on the implementation of each systematic uncertainty used in the study. Each nuisance parameter is varied by $\pm1\sigma$ to produce alternative response matrices and distributions, allowing the likelihood framework to incorporate these effects in the unfolding procedure.

\subsection{Overall Scale Factor}

This systematic represents uncertainties affecting the overall normalisation of both signal and background processes independently. In experimental contexts, this corresponds to integrated luminosity uncertainty, trigger efficiency uncertainties, or theoretical cross-section uncertainties that scale entire distributions without modifying their shapes.

The nominal event weights for each signal and background are multiplied by $\mathrm{NP1}_\mathrm{value} \times (1 \pm\delta)$ where $\delta$ represents the fractional uncertainty. For the studies presented, a nominal value of $\mathrm{NP1}_\mathrm{value}=1$ is used. This variation affects the truth histogram, reconstructed histogram, and response matrix uniformly meaning that all bin contents scale by the same factor while preserving relative bin-to-bin ratios. Critically, this systematic maintains the shape of distributions while shifting their absolute normalizations, representing a fully correlated uncertainty across all bins.

\begin{figure}[!ht]
\includegraphics[width=.9\textwidth]{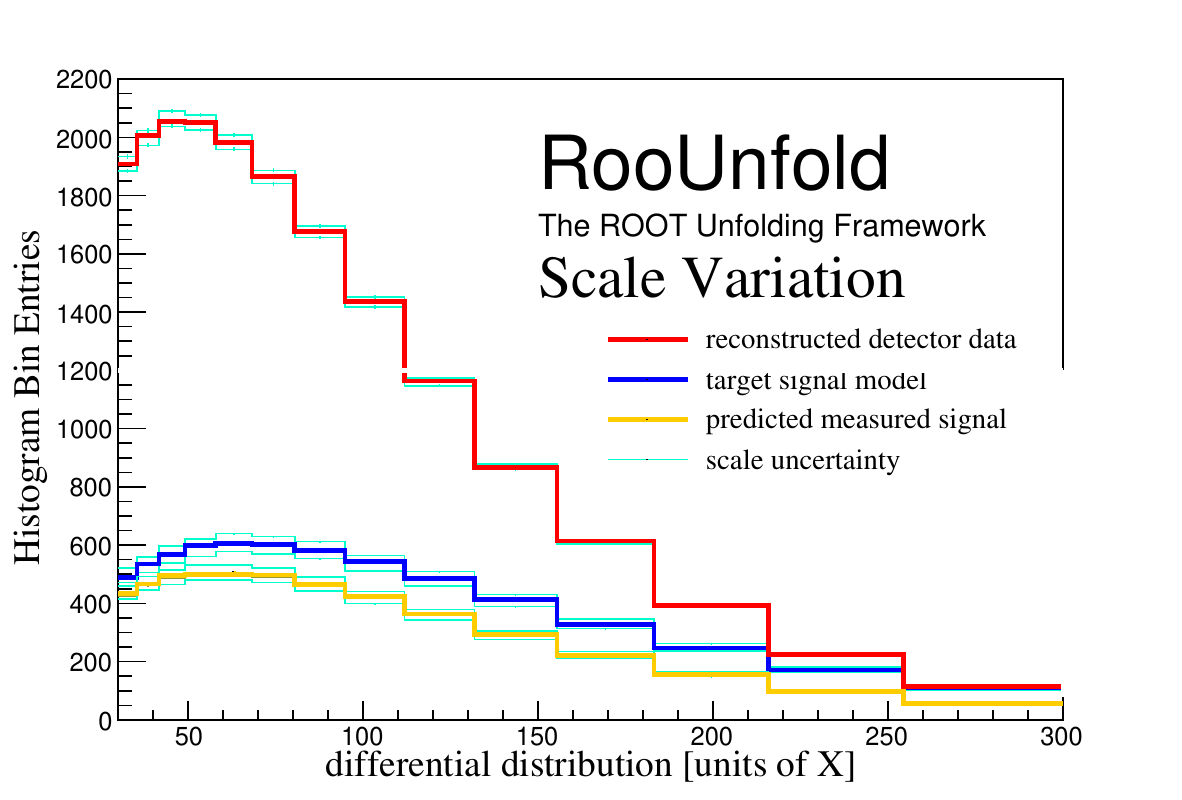}
\caption{The systematic variation of scale factors.}
\label{fig:syst_scale}
\end{figure}

The $\pm1\sigma$ variations produce parallel shifts in the reconstructed spectrum without changing peak positions or widths. The response matrix elements scale proportionally while the conditional probabilities $\mathrm{P(reco|truth)}$ remain unchanged since both numerator and denominator scale together. The scale variations are shown in Figure \ref{fig:syst_scale}.

\subsection{Detector Resolution (Smearing)}

This systematic models uncertainties in the detector's energy or momentum resolution. In calorimeter-based measurements, this can represent jet energy resolution (JER) uncertainties; in tracking systems, it could correspond to momentum resolution uncertainties. The resolution function determines how much the measured value fluctuates around the true value due to instrumental limitations.

The smearing function applies Gaussian fluctuations to the true values: \\ $x_\mathrm{reco} = x_\mathrm{true} + \mathrm{NP2}_\mathrm{value} \times (1 \pm \delta) \times \mathrm{Gaussian}(0, \sqrt{x_\mathrm{true}})$.\\ The nominal value is $\mathrm{NP2}_\mathrm{value} = 1.0$, and variations of $\pm\delta$ modify the width of the Gaussian smearing kernel. Improved resolution (negative variation) produces reconstructed distributions closer to truth with less migration between bins; degraded resolution (positive variation) increases migration and smearing, broadening peaks and enhancing tails.

\begin{figure}[!ht]
\includegraphics[width=.9\textwidth]{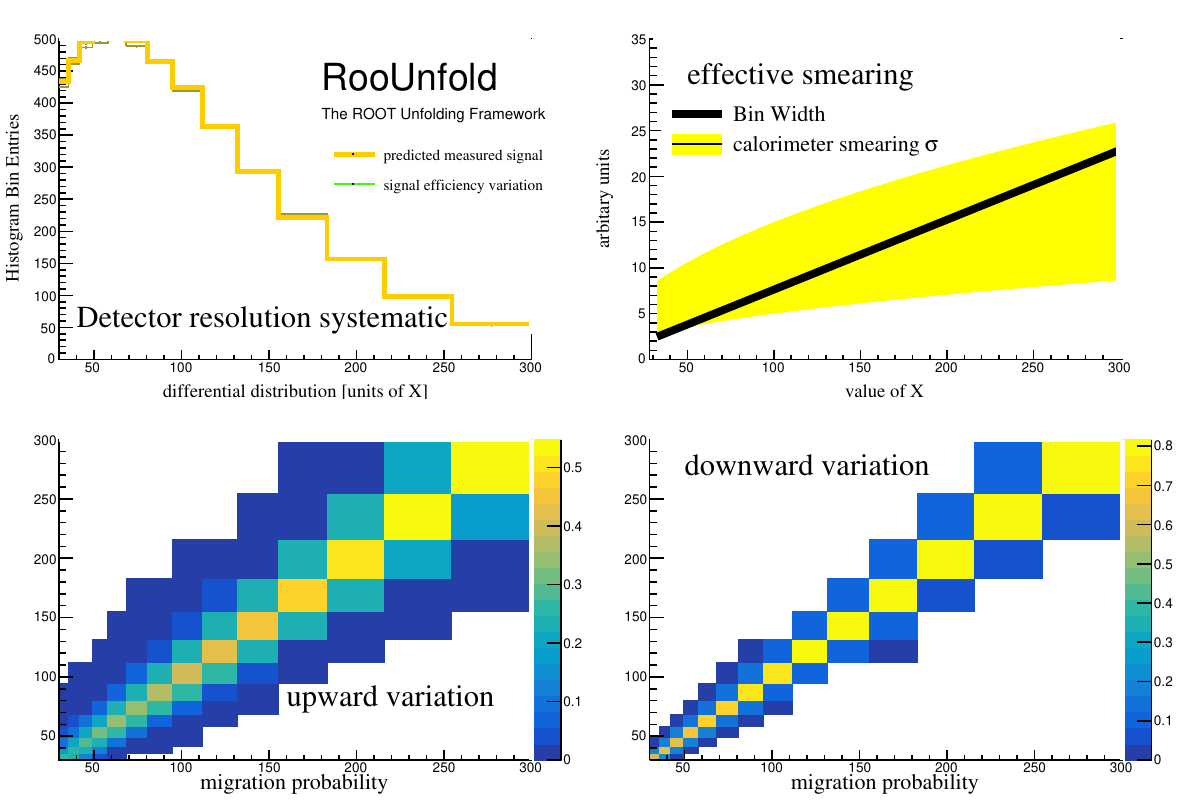}
\caption{The systematic variation due to detector smearing.}
\label{fig:syst_smear}
\end{figure}

Enhanced smearing increases off-diagonal elements in the response matrix, creating more event migration between distant bins. The reconstructed distribution becomes broader and smoother compared to truth. This systematic introduces correlations between neighbouring bins, such that events migrate coherently to adjacent regions rather than randomly redistributing. The smearing variations are shown in Figure \ref{fig:syst_smear}.

\subsection{Measurement Efficiency}

This represents position-dependent or kinematic-dependent detection efficiency. Physical origins include geometric acceptance (particles escaping detector detection), reconstruction efficiency variations (degraded performance at high pseudorapidity or in dense environments), and particle identification efficiency as a function of transverse momentum or other observables. Unlike the scale factor (NP1), this efficiency varies as a function of the measured observable.

An efficiency function $\epsilon(x_\mathrm{true}) = \mathrm{NP3}_\mathrm{value} \times (1 \pm \delta) - |x_\mathrm{true}|/600$ determines event retention probability. The nominal efficiency is $\mathrm{NP3}_\mathrm{value} = 0.95$ at $x=0$, decreasing linearly with $|x_\mathrm{true}|$. Events are randomly rejected based on this efficiency, with higher $|x_\mathrm{true}|$ values having lower acceptance. Varying NP3 shifts the entire efficiency curve up or down while maintaining its functional form, effectively changing the kinematic acceptance window.

\begin{figure}[!ht]
\includegraphics[width=.9\textwidth]{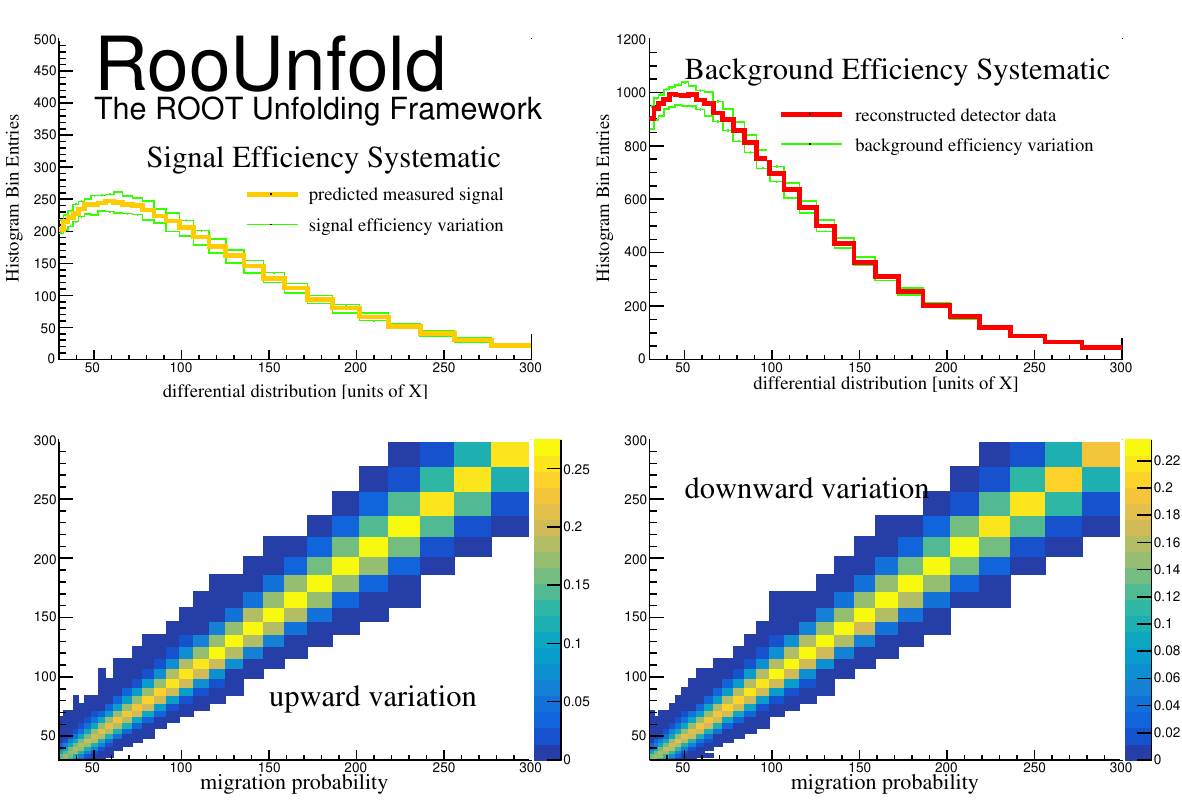}
\caption{The systematic variation efficiency, with the effect on the background shape on the right and the effect on the signal shape on the left. The migration probability for upward and downward variations are show in the bottom panels.}
\label{fig:syst_eff}
\end{figure}

Efficiency variations predominantly affect the extremes of the kinematic range where efficiency gradients are steepest. The response matrix shows non-uniform changes, where bins at high $|x_\mathrm{true}|$ gain or lose events more dramatically than central bins. This systematic creates correlations between the shape of the truth distribution and the observed distribution, as regions with different efficiencies respond differently to the variation. The variation is visualised in Figure \ref{fig:syst_eff}. 

\subsection{Background Shape}

Background processes rarely have perfectly known shapes, but rather they depend on theoretical predictions (which carry PDF, scale, and shower modelling uncertainties) or on extrapolations from control regions (which carry statistical and systematic uncertainties). This nuisance parameter represents shape variations in the background contribution, capturing how background shape mis-modelling would affect the measurement.

The background is modelled as an exponential distribution with slope parameter $\lambda_\mathrm{bkg}$. The nominal value is $\lambda_\mathrm{bkg} = -0.02$, and systematic variations modify this: $\lambda_\mathrm{bkg} = -0.02 \times (1 \pm \delta)$. More negative slopes produce harder (flatter) spectra; less negative slopes produce softer (steeper) spectra. Entirely new background samples are generated for each variation with the modified slope parameter, producing correlated shape changes across the full kinematic range.

The background shape variation affects only the reconstructed distribution and background contribution, and therefore does not alter the response matrix or truth distribution since those depend only on signal properties. The effect is most pronounced where background constitutes a significant fraction of the total reconstructed yield. After unfolding, any residual background contamination or imperfect background subtraction propagates this uncertainty into the final result. The variation is visualised in Figure \ref{fig:syst_bkg_sig} on the left. 

\subsection{Signal Shape}

Theoretical uncertainties in signal modelling arise from PDF choice, renormalization and factorization scale variations, parton shower and hadronisation model differences (Pythia vs. Herwig), and higher-order correction uncertainties. These affect both the predicted truth distribution and, through the response matrix, the relationship between truth and reconstruction. Unlike background shape uncertainties, signal shape systematics impact every stage of the unfolding process.

The signal process follows an exponential distribution with slope $\lambda_\mathrm{sig} = -0.015$ (nominal). Variations modify this parameter: $\lambda_\mathrm{sig} = -0.015 \times (1 \pm \delta)$. Alternative signal samples are generated with the shifted slope parameter, propagating through the full simulation chain including detector smearing and efficiency effects. This produces correlated variations in the truth histogram, reconstructed signal histogram, and response matrix simultaneously.

\begin{figure}[!ht]
\noindent \begin{centering}
\includegraphics[width=.49\textwidth]{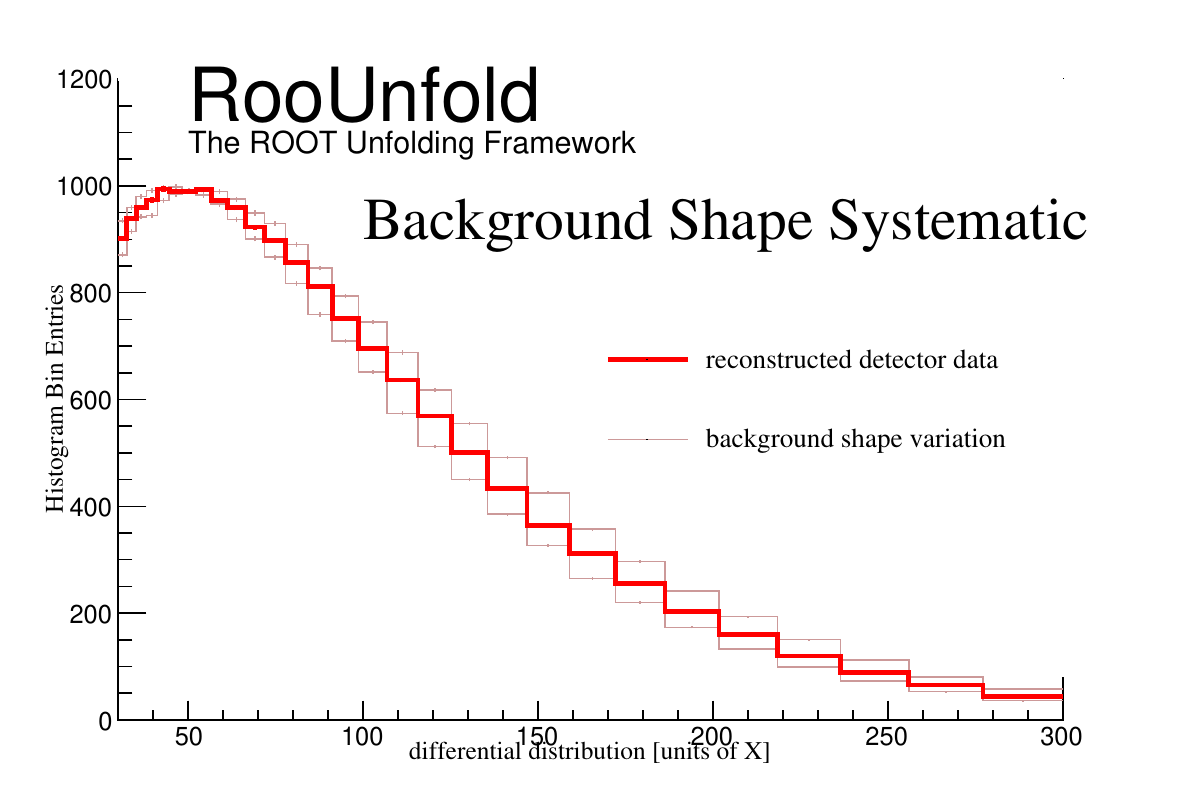}
\includegraphics[width=.49\textwidth]{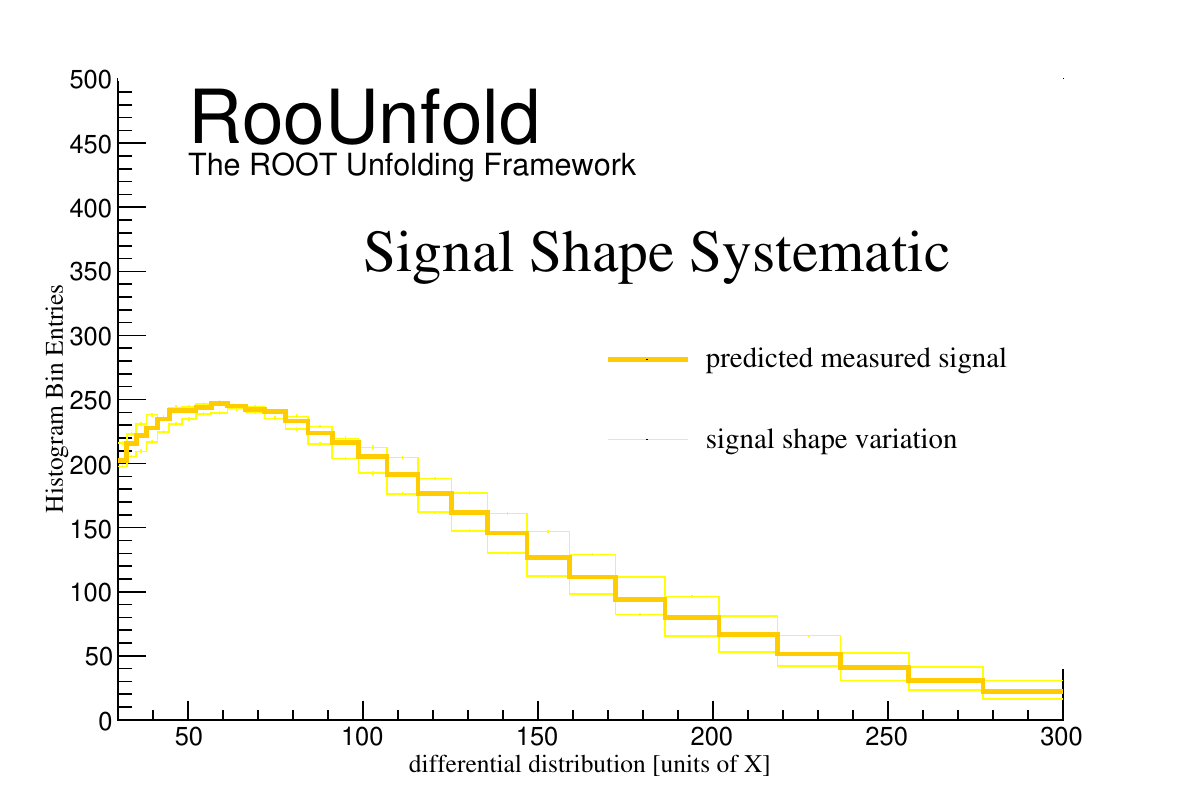}
\par\end{centering}
\caption{Left: The systematic variation on the background shape. Right: The systematic variation on the signal shape.}
\label{fig:syst_bkg_sig}
\end{figure}

Signal shape variations represent the most complex systematic in this study because they affect all components: truth (changing the underlying distribution being measured), response matrix (changing the truth-to-reco mapping), and reconstructed distribution (changing observed yields). 
The variation is visualised in Figure \ref{fig:syst_bkg_sig} on the right. 

\subsection{Total (Luminosity) Scale}

Integrated luminosity uncertainty is one of the most fundamental systematics in collider physics, typically quoted as a 1-5\% uncertainty depending on the experiment and collision system. This uncertainty affects all processes coherently, where both signal and background scale by the same factor, representing uncertainty in the absolute rate calibration. Unlike NP1 (which can vary signal and background independently), this represents a fully correlated scaling across all contributions.

All event weights (signal and background) are scaled uniformly by $1 \pm \delta$. This variation affects total yields while preserving all relative normalizations between signal and background, as well as all shape information. The response matrix scales in absolute normalization but conditional probabilities remain unchanged.

Total luminosity uncertainty enters the unfolding primarily through the constraint on total event yields. Methods that enforce flux conservation or match total yields between data and prediction will show reduced sensitivity to this systematic. The effect should be nearly perfectly correlated across all bins, making it distinguishable from other systematic sources in the covariance matrix.

\subsection{Implementation Details}

All systematic variations follow a common structure in the generation of these variations:
\begin{enumerate}
    \item Baseline distributions are generated with central values of all nuisance parameters
    \item For each systematic source, $\pm 1 \sigma$ variations are produced by modifying the relevant parameter while holding others at nominal values
    \item The probability matrix P(reco$|$truth) is computed for nominal and each variation, accounting for migration between bins
    \item In the likelihood framework, each nuisance parameter $\alpha_i$ has a Gaussian constraint term $\mathcal{N}\tilde{a}_i|\alpha_i,\sigma_i)$ representing the subsidiary measurement, typically with $\sigma_i = 1$ in these studies
\end{enumerate}

The variations are implemented to be realistic while remaining analytically tractable, providing a controlled testing ground for systematic uncertainty treatment in unfolding algorithms.

\end{document}